# DE BROGLIE'S EXACT TRAJECTORIES

Adriano Orefice[1,*], Raffaele Giovanelli[1] and Domenico Ditto[1]

[1]University of Milan,
Department of Agricultural and Environmental Sciences
Production, Landscape, Agroenergy (DiSAA)
*Via G. Celoria 2, Milan, 20133 - Italy*

(*) *Corresponding author*: *adriano.orefice@unimi.it*

**Abstract.** De Broglie's quest for a wave-like approach capable of representing the position of a moving particle, is satisfied, in the case of time-independent external fields, by assuming that each particle runs along the *virtual* trajectories associated, under assigned starting conditions, with a time-independent Schrödinger (or Klein-Gordon) equation.
Just like in classical Dynamics, indeed, the starting conditions determine, from the beginning, a set of *virtual* particle trajectories independent from the very presence of the particles.

## 1 - Introduction

Let us begin our considerations by translating the starting lines of a paper written by Louis de Broglie in 1959 [**1**], 30 years after the Nobel Prize rewarding his foundation of Wave Mechanics:

> In my first works on Wave Mechanics, dating back to 1923, I had clearly perceived that it was necessary, in a general way, to associate with the movement of any corpuscle the propagation of a wave. But the homogeneous wave that I had been led to consider, and that became the wave $\psi$ of the usual wave mechanics, did not seem to me to describe the physical reality (...). Giving no particular prerogative to any point in space, *it was not capable of representing the position of the corpuscle*: we could *suppose* at most, as was very shortly done, that it gave, by its square, the "probability of presence" of the corpuscle in each point.

The present paper aims to satisfy de Broglie's quest, in the case of stationary external fields, without resorting, as we point out in our conclusive Discussion, to any kind of (Bohmian or other) probabilistic, statistical or hydrodynamic approach. As is shown by our numerical examples, our approach is straightforward and self-consistent, and its ultimate test is entrusted to experiment. Our goal is to provide an exact particle dynamics going beyond "the simple illusion of a better understanding" [**2**].

## 2 – The wave-mechanical context

We shall refer, in the following, to the motion of non-interacting point-particles of mass $m$, launched with momentum $p_0$ and energy $E = p_0^2 / 2m$ into a force field deriving from a stationary potential energy $V(\boldsymbol{r})$.
Let us preliminarily remind that, according to *Fermat's* variational principle [**3**], any *ray* of a wave travelling (with wave-vector $\boldsymbol{k}$ ) between two points *A* and *B* shall follow a trajectory satisfying the condition $\delta \int_A^B k\, ds = 0$, where $k = |\boldsymbol{k}|$ and $ds$ is an element of a virtual line connecting *A* and *B*. According to *Maupertuis's* variational principle, on the other hand, any

*material particle* moving between *A* and *B* with momentum $\boldsymbol{p}$ shall follow a trajectory satisfying the analogous condition $\delta\int_A^B p\, ds \equiv 0$, with $p=|\boldsymbol{p}|$.

This formal analogy between *waves* and *matter* suggested to **de Broglie** [**4**] the association of the particle momentum $\boldsymbol{p}(\boldsymbol{r},E)$ with the wave-vector $\boldsymbol{k}(\boldsymbol{r},E)$ of a stationary *"matter wave"* of the general form

$$u(\boldsymbol{r},E) \equiv R(\boldsymbol{r},E)\, e^{\,i\varphi(\boldsymbol{r},E)}\,, \qquad (1)$$

where $R(\boldsymbol{r},E)$ represents the particle distribution over a phase surface $\varphi(\boldsymbol{r},E)= const$ normal to the wave-vector field

$$\boldsymbol{k}(\boldsymbol{r},E) = \boldsymbol{\nabla}\,\varphi(\boldsymbol{r},E)\,, \qquad (2)$$

according to the *Ansatz*

$$\boldsymbol{p}(\boldsymbol{r},E) = \hbar\,\boldsymbol{k}(\boldsymbol{r},E)\,. \qquad (3)$$

Eq.(3), laying the foundation itself of *Wave Mechanics*, implies a spatially periodic feature whose ultimate test, as we wrote in the Introduction, is entrusted to experience**.**

It plays the threefold role of:

1) **predicting a wavelength $\lambda \equiv 2\pi / k = 2\pi\hbar / p$, which was verified by the Davisson-Germer experiments [5, 6] on electron diffraction by a crystal of Nickel,**
2) **addressing the particle momentum $\boldsymbol{p}$ along the field-lines of the wave-vector $\boldsymbol{k}$, and**
3) **suggesting, therefore, that particle trajectories do exist, contrary to widespread different opinions.**

Concerning the choice of the equation describing the "matter wave" modality, we recall that *any Helmholtz equation* (the simplest form of stationary wave equation) is associated with an exact Hamiltonian system of coupled ray-trajectories [**7-12**].

Keeping therefore eqs.(2) and (3) into account we look (according to *Schrödinger's* suggestion [**13**]) for a Helmholtz-like equation whose ray-trajectories reduce, *in the eikonal limit*, to the ones of *Classical Dynamics*. Not surprisingly, such an equation assumes the form of the well-known time-independent Schrödinger equation

$$\nabla^2 u(\boldsymbol{r},E) + \frac{2m}{\hbar^2}[E - V(\boldsymbol{r})]\; u(\boldsymbol{r},E) = 0\,. \qquad (4)$$

Looking, in fact, for solutions of the general form (1) it is easily seen that, *in the eikonal limit* [**3, 11**] of eq.(4), we have $k^2 \simeq \frac{2m}{\hbar^2}\left[E - V(\boldsymbol{r})\right]$, so that, making use of eqs. (2) and (3), we have $p^2 \equiv (\hbar\, k)^2 \simeq 2\, m\; (E - V(\boldsymbol{r}))$, in agreement with *Classical Dynamics.*

By introducing the expression (1) into eq.(4), and separating real from imaginary parts [**7**], the particle motion turns out to occur along the self-contained Hamiltonian system of coupled trajectories

$$\begin{cases} \dfrac{d\boldsymbol{r}}{dt} = \dfrac{\partial H}{\partial \boldsymbol{p}} \equiv \dfrac{\boldsymbol{p}}{m} & (5.1) \\[2ex] \dfrac{d\boldsymbol{p}}{dt} = -\dfrac{\partial H}{\partial \boldsymbol{r}} \equiv -\boldsymbol{\nabla}\;[V(\boldsymbol{r}) + W(\boldsymbol{r},E)] & (5.2) \\[2ex] \boldsymbol{\nabla} \bullet (R^2\,\boldsymbol{p}) = 0 & (5.3) \end{cases} \qquad (5)$$

where $$W(\boldsymbol{r},E) = -\frac{\hbar^2}{2m}\frac{\nabla^2 R(\boldsymbol{r},E)}{R(\boldsymbol{r},E)} \tag{6}$$

and $$H(\boldsymbol{r},\boldsymbol{p},E) \equiv \frac{p^2}{2m} + W(\boldsymbol{r},E) + V(\boldsymbol{r}) . \tag{7}$$

The construction, now, of the "matter wave" trajectories provided by eqs. (5) all over the region of particle motion requires the **preliminary assignment of a wave-mechanical context**, consisting of the **launching momentum and position** of the particles: whatever the experimental set-up may be, the specification of this assignment is the basic condition for the interpretation of the experimental results.
According, therefore, to eqs.(1), (2) and (3), the time-integration of the equation system (5) is started by assigning the particle distribution $R_0(\boldsymbol{r},E)$ over a launching wave-front $\varphi_0(\boldsymbol{r},E) = const$, where $\boldsymbol{p}_0(\boldsymbol{r},E) = \hbar\nabla\varphi_0(\boldsymbol{r},E)$, and $p_0 = \sqrt{2mE}$.
Eq.(5.3) (expressing the constancy of the flux of the vector field $R^2\boldsymbol{p}$ along any tube formed by the ray-trajectories) shall provide, at each time-step of the particle motion, the values of the amplitude distribution $R(\boldsymbol{r},E)$, and therefore of the Wave Potential $W(\boldsymbol{r},E)$, on the *next* wave-front $\varphi(\boldsymbol{r},E) = const$.
The **trajectory-coupling** “Wave Potential” term $W(\boldsymbol{r},E)$ is the one and only cause of any (stationary) wave-like process, such as diffraction and/or interference, and in its absence the Hamiltonian system (5) is seen to reduce, as expected, to standard Classical Dynamics. A further property of $W(\boldsymbol{r},E)$ is to act perpendicularly to the particle momentum [**7**], leaving therefore its energy unchanged.
Notice that the stationary, energy-dependent Wave Potential $W(\boldsymbol{r},E)$ has little to share (in spite of its formal analogy) with Bohm’s almost intractable "Quantum Potential" $Q(\boldsymbol{r}, t)$ [**14**], mixing together the whole set of energy eigen-values of eq.(4).

An impressive synthesis of matter, energy and waves is obtained, generalizing the case of eqs. (4) and (5), in the *relativistic case*, where eq.(4) is replaced, for particles of rest mass $m_0$, by a suitable (Helmholtz-like) Klein-Gordon equation [**9-11**], associated with a coupled trajectory system (holding even in case of vanishing rest mass) of the form:

$$\begin{cases} \dfrac{d\boldsymbol{r}}{dt} = \dfrac{\partial H}{\partial \boldsymbol{p}} \equiv \dfrac{c^2\,\boldsymbol{p}}{E - V(\boldsymbol{r})} & (8.1) \\ \dfrac{d\boldsymbol{p}}{dt} = -\dfrac{\partial H}{\partial \boldsymbol{r}} \equiv -\nabla V(\boldsymbol{r}) - \dfrac{E}{E - V(\boldsymbol{r})}\nabla W(\boldsymbol{r},E) & (8.2) \\ \nabla \bullet (R^2\,\boldsymbol{p}) = 0 & (8.3) \end{cases} \tag{8}$$

with $$W = -\frac{\hbar^2 c^2}{2E}\frac{\nabla^2 R}{R} \tag{9}$$

and $$H \equiv V + \sqrt{(pc)^2 + (m_0 c^2)^2 + 2EW} . \tag{10}$$

It may be observed that eq.(8.1) is a particular case of the relativistic "*guidance law*" which expresses, according to de Broglie’s (*unfortunately incomplete*) "Double Solution Theory" [**15**], the velocity of “*minute singular regions*" (representing physical particles) non-linearly included in an *objective physical wave*, associated with a *subjective wave* of statistical significance.

*Contrary to the popular narrative*, but in agreement both with the Davisson-Germer experiment [**5**] and with the "ontologic parsimony" expressed by Ockham's Razor [**16**], we may conclude that the wave-mechanical "duality" *does not predict a physical "matter wave" travelling with the particle*. It predicts, starting from an assigned set of phase and amplitude distributions $\varphi_0(\boldsymbol{r},E)$ and $R_0(\boldsymbol{r},E)$ (the **wave-mechanical context**), a *stationary virtual system* of trajectories, coupled by the stationary Wave Potential (9) and provided by the Hamiltonian systems (5) or (8). Just like in Classical Mechanics, indeed, the starting conditions determine a *virtual* set of trajectories (and time-tables) independent from the number, and even from the very presence, of moving particles. Once the launching conditions are applied, the time-integration of the relevant trajectory system provides step by step, without resorting to any kind of probabilistic or hydrodynamic interpretation, **the solution of de Broglie's problem of representing the exact position of any "corpuscle"** all along its motion. We shall obtain, in fact, a stationary pattern of *virtual* trajectories stemming from the launching wave-front and mutually coupled, at each point of space, by the local value of the Wave Potential. Both the exact trajectory and time-table of any single particle shall be simply picked up from the overall display of numerical results, starting from its launching position. In the absence of wave-mechanical coupling the trajectory pattern and motion law would clearly reduce, as already observed, to Classical Mechanics.

Limiting ourselves, here, to the *non-relativistic* system (5), let us expound the relevant physics in detail, by means of a few simple examples of wave-mechanical contexts, **borrowed from our own Ref. [11].** Assuming, for simplicity sake, a geometry limited to the $(x,z)-$plane, we consider, to begin with, the *diffraction* of particles launched along the $z$-axis, in the absence of any external field $V(\boldsymbol{r})$, through a slit of half-width $w_0$ (centered at $x=0,\ z=0$) practiced on a screen placed along the *x*-axis. Any particle launched with momentum components $p_x(t=0)=0$ and $p_z(t=0)=p_0=\sqrt{2mE}$ shall be guided by a matter wave with launching wavelength $\lambda_0=2\pi\hbar/\sqrt{2mE}$. In the absence of external fields, and under the rule of the *energy-preserving* Wave Potential alone, we'll have $\lambda=\lambda_0$ all along the particle motion. The numerical solution of the Hamiltonian system (5) shall provide both a stationary pattern of *coupled* trajectories and the particle *time-table* along these *"rails"*.

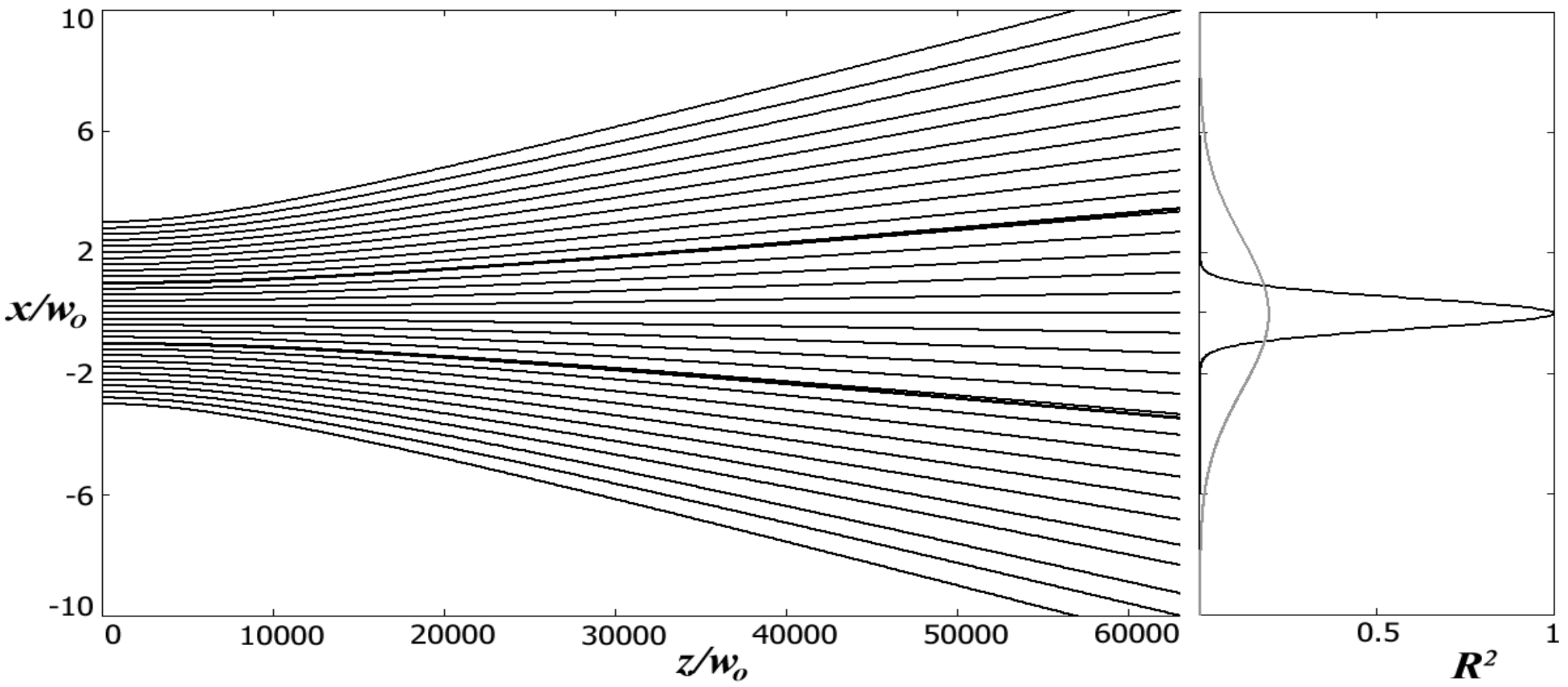

**Fig.1 – Particle trajectories in the case of a Gaussian launching distribution.**

**Fig.1** refers to the case of particles launched along the *z-axis* with a transverse amplitude distribution of the *Gaussian* form [**10**] $R(x;z=0)\propto exp(-x^2/w_0^2)$, with $\lambda_0/w_0\simeq 10^{-4}$.

The full virtual *trajectory pattern* is plotted on the *left-hand* side of the figure, in terms of the dimensionless co-ordinates $x/w_0$ and $z/w_0$. The *right-hand* side presents, in its turn, the initial and final intensity ($R^2$) distribution. The *two heavy lines* (representing particle trajectories starting from the so-called *"waist" positions* $x/w_0 = \pm 1$ ) turn out to be in *excellent numerical agreement* with their well-known "paraxial" [**17**] analytical expression

$$\frac{x}{w_0} = \pm\sqrt{1+\left(\frac{\lambda_0 z}{\pi w_0^2}\right)^2} \,, \qquad (11)$$

thus providing a test of our approach.

The *dramatic dependence* of the particle trajectories on the *wave-mechanical context* (i.e. on the launching assignment of $\varphi_0(\boldsymbol{r},E)$, $R_0(\boldsymbol{r},E)$ and $\lambda_0/w_0$ ) is clearly shown in the simple example of **Fig.2,** where particles are launched with a bell-shaped *non-Gaussian* amplitude distribution [**10**]. While in the previous case, as predicted by the standard optical diffraction theory [**18**], the diffracted wave distribution preserves its Gaussian form, the same doesn't hold in the non-Gaussian case, where diffraction "fringes" are seen to appear.

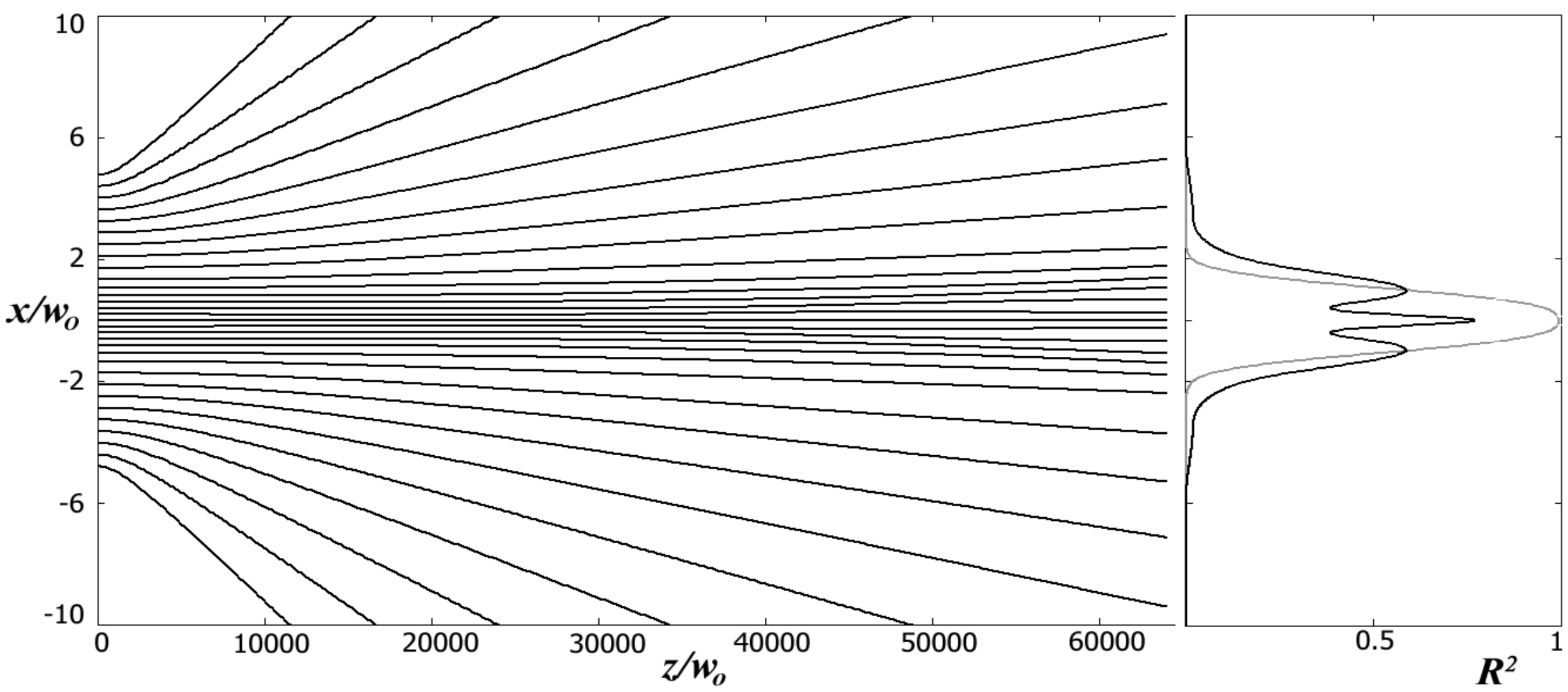

**Fig.2 - Particle trajectories in the case of a non-Gaussian launching distribution.**

We finally present, in **Fig.3,** the case of a *Gaussian* particle distribution launched along the *z-axis* into an external potential field $V(x,z) = E(1-n^2)$, representing, for $n(x,z) = 1 + a\, exp\left[-(\frac{x}{b})^2 - (\frac{z-c}{d})^2\right]$ and with a suitable choice of the constants $a,\ b,\ c,\ d,$ a lens-like focalizing structure [**19**]. The *point-like focus* obtained in the Geometrical Optics limit (dashed trajectories), by dropping the Wave Potential term $W(\boldsymbol{r},E)$ from the equation system (5), is seen to be replaced by a *finite focal waist* when the diffractive role of the Wave Potential is duly taken into account (continuous trajectories).

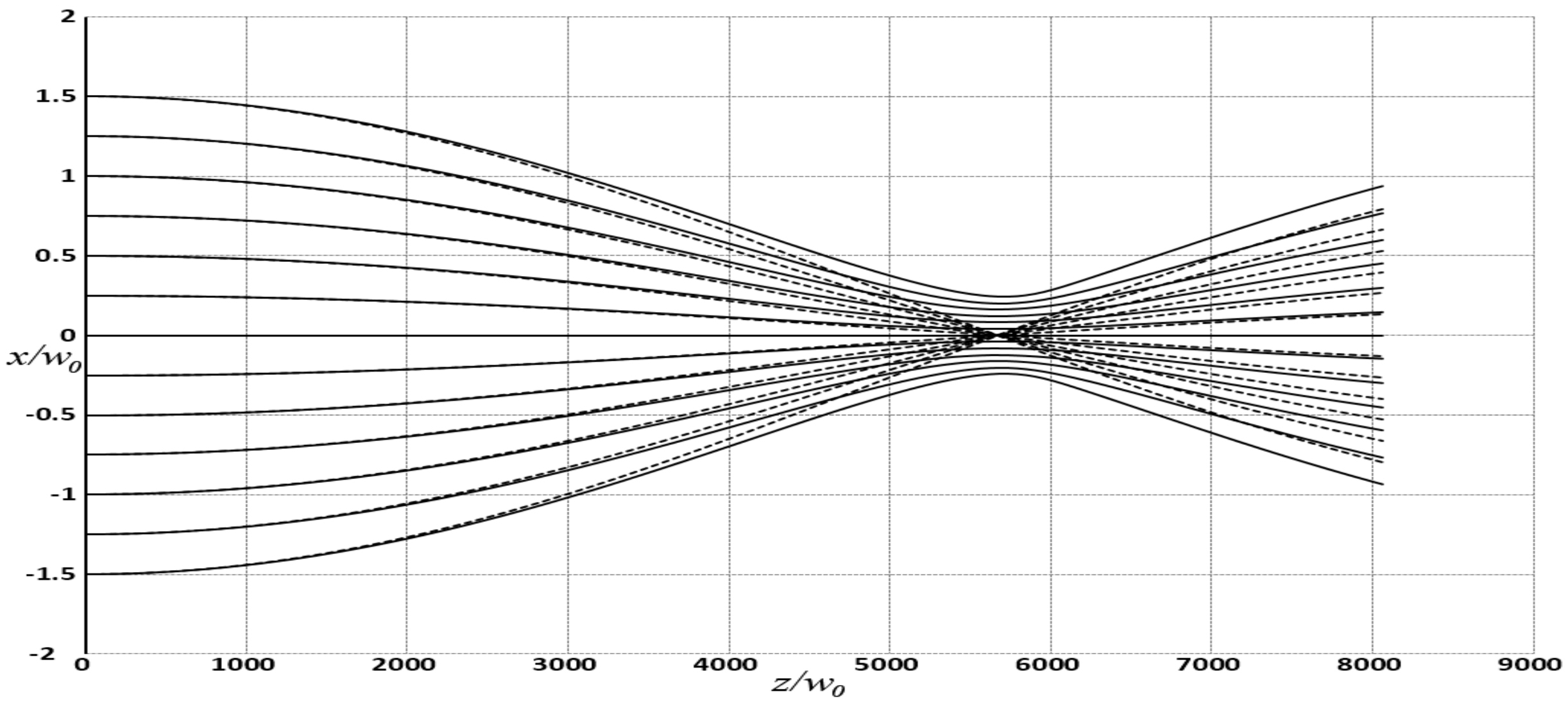

**Fig.3 - Particle trajectories in a lens-like potential.**

## 3 – Discussion: Wave-Mechanics versus Standard Quantum Mechanics

The present paper stresses the basic role of the *time-independent* Schrödinger equation (4) in the wave mechanical context. In fact,

1) besides providing the main standard items of the whole quantum theory, such as energy eigen-values, harmonic oscillators and atomic energy levels [**6**],
2) eq.(4) lends itself to describe, thanks to its Helmholtz-like nature, the particle motion along a virtual pattern (5) of coupled trajectories, determined by the launching conditions, under the diffractive and energy preserving action of the Wave Potential. Moreover,
3) thanks to its energy-dependence, it presents the Wave-Mechanical Dynamics as a direct extension of Classical Dynamics and of its *virtual* trajectories, to which it includes a trajectory coupling and energy-dependent force term, $\nabla W(\boldsymbol{r},E)$, of wave-like origin.

And what about the *time-dependent* Schrödinger equation? Under Planck's condition

$$E=\hbar\omega \,, \tag{12}$$

let us refer to a physical case involving a discrete set of eigen-energies $E_n$, eigen-frequencies $\omega_n = E_n / \hbar$ and eigen-solutions $u_n(\boldsymbol{r})$ of eq.(4) , and let us "complete" the matter wave (1) in the form

$$\psi_n(\boldsymbol{r},\, t) = u_n(\boldsymbol{r})\, e^{-i\,\omega_n t} \equiv u_n(\boldsymbol{r})\, e^{-i\, E_n t/\hbar} \tag{13}$$

Any linear superposition, now, of the form

$$\psi(\boldsymbol{r},t) \;=\; \sum_n c_n\, \psi_n(\boldsymbol{r},t)\,, \tag{14}$$

with arbitrary constant coefficients $c_n$, turns out to be a solution of the *time-dependent* Schrödinger equation

$$i\,\hbar\,\frac{\partial\psi}{\partial t} = -\,\frac{\hbar^2}{2m}\nabla^2\,\psi + V(\boldsymbol{r})\,\psi\,, \tag{15}$$

directly obtained from eqs. (1), (4) and (13): *in itself, NOT a standard-looking wave equation*. The insertion of the function (14) into eq.(15) puts it into the form

$$\sum_n c_n \, e^{-i\frac{E_n}{\hbar}t} \left\{ \nabla^2 u_n(\boldsymbol{r}) + \frac{2m}{\hbar^2}[E_n - V(\boldsymbol{r})]\, u_n(\boldsymbol{r}) \right\} = 0 \,, \qquad (16)$$

where each term of the sum is of the form (4). Eq. (16) provides the *closest approach* between the *time-independent* Schrödinger eq.(4) and the *time-dependent* eq.(15), involving the role of the full assembly of the particle states.

*Two different routes* are usually followed in the way of dealing with eqs.(15) and (16):

**1) Standard Quantum Mechanics** (SQM) is centered on eq.(15), which is taken as a *Fundamental Postulate* (see Ref.[**20**], Chapt.III, page 52). The function (14), with the help of eq.(16), is interpreted as carrying the most complete physical information, ranging over the full set of particle eigen-energies, and collapsing into one of them, in case of observation, according to the (duly normalized) probability $|c_n|^2$ [**21**].

A "probability current density" is obtained in the form

$$\boldsymbol{J} \equiv \frac{\hbar}{2mi}(\psi^* \nabla\psi - \psi \nabla\psi^*), \qquad (17)$$

where $\psi\,\psi^* \equiv |\psi|^2 \equiv R^2$, and *no definite particle trajectory* (such as eqs.(5)) is admitted, in an odd-looking "scrambling of ontic and epistemic contextualities" **[22, 23]**.

**2) The Bohmian approach** [**14, 24**] assumes, in turn, a (wave-packet) "guidance velocity" provided by a mixture of *dynamics* and *hydro-dynamics* of the form

$$\frac{d\,\boldsymbol{r}(t)}{dt} \equiv \boldsymbol{J}\,/\,R^2 \equiv \frac{\hbar}{2mi}\,\frac{\psi^* \nabla\psi - \psi\nabla\psi^*}{\psi\,\psi^*} \;, \qquad (18)$$

to be evaluated in parallel with the numerical solution, $\psi(\boldsymbol{r},t)$, of eq.(15), and addressing the particle motion along the flux lines "*along which the probability density is transported*" [**24**]. The term $\frac{d\boldsymbol{r}}{dt}$ represents the velocity of a (hopefully and permanently) narrow wave-packet centered at $\boldsymbol{r}(t)$: a guess which may give rise, in general, to analytical and computational troubles [**25**] . We remind, in this connection, that the original formulation of the Bohmian theory [**14**] involved a *time-dependent* (and *energy-independent*) "Quantum Potential" function $Q(\boldsymbol{r},t) = -\frac{\hbar^2}{2m}\frac{\nabla^2\psi(\boldsymbol{r},t)}{\psi(\boldsymbol{r},t)}$, formally similar to (but *utterly different from*) the "Wave Potential" function $W(\boldsymbol{r},E)$ of our eq.(6), whose time-independence and monochromaticity makes it apt to describe diffraction and/or interference processes. The exploitation of Bohm's Quantum Potential, however, didn't always lead to reliable numerical procedures [**24-26].**

Both in SQM and in the Bohmian approach, in conclusion, the *exact, deterministic dynamics* allowed, for each particle, by eq.(4) is replaced by a *probability flow* [**25**] based on eq.(15), and the role, experimentally well-established by Davisson and Germer [**5**], of de Broglie's monochromatic "matter waves" is, to say the least, overshadowed, but certainly not contradicted.

Both SQM and Bohmian Mechanics aim indeed, from the very beginning, to a description based on a full set of particle eigen-energies, telling us, with respect to de Broglie's Wave Mechanics, *a quite different story.*

Let us also notice that:

**-** the *assumption* of de Broglie's Ansatz in the (popular but incomplete) **scalar** form $\lambda = 2\pi\hbar / p$ (rather than in the complete **vectorial** form (3), directly suggesting the existence of particle trajectories), and
**-** the *assumption* of the (strongly counter-intuitive) eq. (15) as a **Fundamental Postulate**, prevailing on eq.(4), which is however, and ought to remain, its basic tool,
have contributed to the widespread reputation of Quantum Mechanics as a scientific mystery.
Indeed, "*there remain several features of Quantum Mechanics that resist an intuitive explanation. It is an open question whether unexpected future developments will produce an explanation of these features, or they will just become familiar by getting used to it. After all, "understanding" is a human mental state conditioned by our brain that has reached its state after a few million years of evolution, but might not necessarily be adequate for the microscopic world*" [**27**].